\documentclass{tMPH2e}
\usepackage[UKenglish]{babel}
\usepackage[utf8]{inputenc}
\usepackage{amsmath,amssymb}
\usepackage{mathtools}
\usepackage{color}
\usepackage[normalem]{ulem}

\newcommand*{\mat}[1]{\boldsymbol{#1}}
\newcommand*{\coord}[1]{\mathbf{#1}}
\newcommand*{\vecr}{\coord{r}}

\newcommand*{\vecx}{\coord{x}}

\newcommand*{\binteg}[3]{\int^{\mathrlap{#3}}_{\mathrlap{#2}}\ud{#1}\,}

\DeclareMathOperator{\Laplace}{\mathcal{L}}

\DeclarePairedDelimiter{\abs}{\lvert}{\rvert}

\DeclarePairedDelimiterX\braket[2]{\langle}{\rangle}{#1\delimsize\vert#2}
\DeclarePairedDelimiterX\brakket[3]{\langle}{\rangle}{#1\delimsize\vert#2\delimsize\vert#3}
\DeclarePairedDelimiterX\ket[1]{\lvert}{\rangle}{#1}
\DeclarePairedDelimiterX\bra[1]{\langle}{\rvert}{#1}

\newcommand*{\e}{\mathrm{e}}
\newcommand*{\im}{\textrm{i}}
\newcommand*{\isDefinedAs}{\coloneqq}
\newcommand*{\half}{\frac{1}{2}}
\newcommand*{\ud}{\mathrm{d}}

\title{Invertibility of the retarded response functions for initial mixed states: application to one-body reduced density matrix functional theory}
\author{K.J.H. Giesbertz$^{\ast}$%
\thanks{$^{\ast}$Email: k.j.h.giesbertz@vu.nl \vspace{6pt}} \\\vspace{6pt}
\textit{Section of Theoretical Chemistry, Faculty of Exact Sciences, VU University, De Boelelaan 1083, 1081 HV Amsterdam, The Netherlands}}

\begin{document}

\maketitle

\begin{abstract}
In [J.~Chem.\ Phys.~\textbf{143}, 054102 (2015)] I have derived conditions to characterize the kernel of the retarded response function, under the assumption that the initial state is a ground state.
In this article I demonstrate its generalization to mixed states (ensembles). To make the proof work, the weights in the ensemble need to be decreasing for increasing energies of the pure states from which the mixed state is constructed. The resulting conditions are not easy to verify, but under the additional assumptions that the ensemble weights are directly related to the energies and that the full spectrum of the Hamiltonian participates in the ensemble, it is shown that potentials only belong to the kernel of the retarded response function if they commute with the initial Hamiltonian. These additional assumptions are valid for thermodynamic ensembles, which makes this result also physically relevant. The conditions on the potentials for the thermodynamic ensembles are much stronger than in the pure state (zero temperature) case, leading to a much less involved kernel when the conditions are applied to the retarded one-body reduced density matrix response function.
\end{abstract}

\section{Introduction}
In a recent article~\cite{Giesbertz2015} I generalized the invertibility theorem by Van Leeuwen for the retarded (causal) density response function~\cite{Leeuwen2001} to the retarded one-body reduced density matrix (1RDM) response function. This is an important step forward for the formal foundations of linear-response time-dependent 1RDM functional theory, since a Runge--Gross type of proof is lacking.

This generalization is achieved by cutting the proof by Van Leeuwen into two parts. The first part of the proof can be generalized to completely arbitrary retarded response functions, under following two assumptions. The initial (reference) state should be a ground state and the perturbing potential should be Laplace transformable in time. This first part yields a necessary condition which potentials need to satisfy to yield a zero response, i.e.\ to belong to the kernel of the retarded response function. For non-degenerate ground states this condition is also sufficient, but in the degenerate case an additional sufficiency condition needs to be checked. These conditions depend on the operators for which the retarded response function is considered, e.g.\ dipoles, magnetization and spin-density. Hence, the second step of the proof consists of inserting the operators under consideration into the conditions and to check to which extend they are satisfied.

In Ref.~\cite{Giesbertz2015} I have explicitly worked out these conditions for the reatarded density response function and the 1RDM response function. For non-degenerate ground states the same result as Van Leeuwen was recovered: the density response function is invertible up to a constant shift in the potential. Additionally, I have worked out the case for degenerate response functions, thereby extending the validity of Van Leeuwen's theorem to degenerate ground states. Not surprisingly, the same result was obtained: the density response function is invertible up to a constant. From the Runge--Gross theorem we already know that this holds for analytic potentials~\cite{RungeGross1984} and even for more general potentials by the work of Ruggenthaler, Penz and Van Leeuwen~\cite{RuggenthalerLeeuwen2011, RuggenthalerGiesbertzPenz2012, RuggenthalerPenzLeeuwen2015}.

More interesting results were obtained by working out the invertibility conditions for the 1RDM response function. All symmetry operators which can be expressed as one-body operators belong to the kernel of the 1RDM response function. So apart from the constant potential, also spin-projection operators, $\hat{\mat{S}}$, belong to the kernel of the 1RDM operator if the Hamiltonian is spin-independent. Likewise, the (angular) momentum operators, $\hat{\mat{L}}$ and\slash{}or $-\im\nabla$, are also candidates if the system is rotationally and\slash{}or translationally invariant. Discrete symmetry operators can only be expressed as one-body operators for non-interacting systems. In most interacting systems, the interaction prevents such a simple expression and more-than-one-body operators are needed to express discrete symmetry operators. Therefore, discrete symmetries are typically of no concern in interacting systems, e.g.\ the Coulomb interaction.

Not only potentials related to the symmetry of the system possibly lead to no response. Also somewhat `pathological' cases need to be considered which are more related to the structure of the ground state than to any symmetry of the system. These potentials can be revealed by working in the natural orbital (NO) representation. The NOs are defined as the eigenfunctions of the 1RDM and the corresponding eigenvalues are called (natural) occupation numbers, $n_k$. By expanding the ground state in Slater determinants constructed from the NO basis, one can easily find that potentials coupling only completely unoccupied NOs, $n_k = 0$, are also part of the kernel of the 1RDM response function. Likewise in the case of fermions, the potentials coupling the fully occupied NOs, $n_k = 1$, also yield no response in the 1RDM. This result does not come as a surprise, since this freedom to make unitary transformations among the (un)occupied orbitals is often exploited in Hartree--Fock and Kohn--Sham. For systems with a Coulomb interaction it is actually highly unlikely that fully occupied or completely empty NOs exist~\cite{MorrisonZhouParr1993, Kimball1975, Friesecke2003, GiesbertzLeeuwen2013a, GiesbertzLeeuwen2013b, Giesbertz2014}. In the case of non-interacting systems, one can actually show that potentials coupling degenerate NOs, $n_k = n_l$, also do not lead to a response in the 1RDM. The last case which needs to be mentioned, is the two-electron wave function. Due to its special structure in the NO representation (the NOs only occur once in pairs~\cite{LowdinShull1956, Kutzelnigg1963, CioslowskiPernalZiesche2002, PhD-Giesbertz2010, RappBricsBauer2014}), special potentials which couple these paired NOs or degenerate pairs also belong to the kernel of the 1RDM response function.

In this article I will consider a further extension of the invertibility proof from pure states to ensembles as initial condition. This allows us to release the constraint of using a ground state as initial state, though we need to demand that the weights of lower lying states are always larger than the weights of higher lying states in the ensemble. Two necessary and sufficient conditions for the potentials in the kernel of the retarded response function are derived for such ensembles. Unfortunately,  it is hard to make any general statements based on these conditions. If, however, the natural assumption is made that the ensemble weights only depend on the energy, only one of these conditions remains. In the case of the (macro)canonical ensemble, the remaining condition further simplifies to the condition that only potentials which commute with the Hamiltonian belong to the kernel of the response function. Only the symmetry related potentials in the kernel of the 1RDM response function therefore remain at finite temperature.

The paper has a comparable structure to Ref.~\cite{Giesbertz2015}. In Sec.~\ref{sec:necCond} I derive a necessary condition and in Sec.~\ref{sec:sufCond} an additional sufficiency condition and discuss how it is trivially satisfied if the weights of the ensemble are purely energy related. In Sec.~\ref{sec:thermoEnsembles} I discuss the implication for the canonical and grand canonical ensembles and show how the kernel of the 1RDM response function is cleaned up when using a finite temperature formalism. In Sec.~\ref{sec:timeINdependent} I discuss how these results relate to the time-\emph{in}dependent 1RDM response function and conclude in the last section.

\section{A necessary condition for potentials in the kernel of the retarded response function}
\label{sec:necCond}
The invertibility proof for general retarded response functions proceeds in a very analogous manner as for initial ground states~\cite{Giesbertz2015}. For the sake of clarity, a full exposition of the proof is useful, since at some key-points new conditions need to be imposed for the proof to work and the ensemble weights need to be included. We will start by deriving a necessary condition which potentials in the kernel of the response function need to satisfy. In the next section we will consider to which extend this condition is sufficient.

We start from a general set of self-adjoint operators or operator densities, $\{\hat{Q}_i\}$, where the index $i$ can perfectly be some multi-index. In the case of operator densities, the index $i$ will be a continuous label and the summations later on should be interpreted as integrals. A typical example would be the density operator, $\hat{n}(\vecr)$. Of course, a mixture of is also perfectly allowed, e.g.\ the spin-density $\hat{n}(\vecx)$ where $\vecx \isDefinedAs \vecr\sigma$ is a combined space and spin coordinate.

Now we consider perturbations to the system by these operators with strengths $\delta v_j(t)$, so we consider the following perturbation to the original time-independent Hamiltonian, $\hat{H}_0$,
\begin{align*}
\delta \hat{V}(t) = \sum_j\hat{Q}_j\delta v_j(t) \, .
\end{align*}
The linear response in the expectation values of the same set of operators can be expressed as~\cite{FetterWalecka1971, PhD-Faassen2005, PhD-Giesbertz2010}
\begin{align*}
\delta Q_i(t) = \sum_j\binteg{t'}{0}{t} \chi_{ij}(t-t')\delta v_j(t') \, ,
\end{align*}
where $\chi_{ij}(t-t')$ is the retarded linear response function which is readily generalized to initial ensembles as
\begin{align*}
\chi_{ij}(t-t') \isDefinedAs -\im\theta(t - t')\sum_Lw_L \brakket{\Psi_L}{[\hat{Q}_{H_0,i}(t),\hat{Q}_{H_0,j}(t')]}{\Psi_L} \, ,
\end{align*}
where $w_L \geq 0$ are the weights of the the states $\ket{\Psi_L}$ in the ensemble and should sum to one.
In the definition of the response function, I have used the interaction picture, so $\hat{Q}_{H_0,i}(t) \isDefinedAs \e^{\im\hat{H}_0t}\hat{Q}_i\e^{-\im\hat{H}_0t}$. In other words, the operators are in the Heisenberg representation with respect to the unperturbed Hamiltonian, $H_0$. To make the response function causal (retarded), the definition also includes a Heaviside function, which is defined as
\begin{align*}
\theta(x) \isDefinedAs \begin{cases}
1 &\text{for $x > 0$} \\
0 &\text{for $x < 0$} \, .
\end{cases}
\end{align*}
The retarded response function for ensembles can be expressed in its Lehmann representation by inserting a complete set of eigenstates of the unperturbed Hamiltonian, $\hat{H}_0$, which gives
\begin{align}\label{eq:Lehmann}
\chi_{ij}(t-t') &= \im\theta(t-t')\sum_{KL}w_L\,\e^{\im\Omega_{KL}(t-t')}q_j^{LK}q_i^{KL} + \text{c.c.} \, ,
\end{align}
where $\Omega_{KL} \isDefinedAs E_K - E_L$ are (de-)excitation energies of $\hat{H}_0$ and $q^{KL}_i \isDefinedAs \brakket{\Psi_K}{\hat{Q}_i}{\Psi_L}$. With the help of the Lehmann representation of the response function we can write the perturbation in the expectation values as
\begin{align*}
\delta Q_i(t) = \im\sum_{KL}w_L\,\binteg{t'}{0}{t}
q_i^{KL}\,a_{LK}(t')\e^{\im\Omega_{KL}(t-t')} + \text{c.c.} \, ,
\end{align*}
where
\begin{align*}
a_{KL}(t) \isDefinedAs \sum_j q_j^{KL}\delta v_j(t) \, .
\end{align*}
The time-integral has the form of a convolution product, which is readily deconvoluted by taking the Laplace transform
\begin{align*}
\Laplace[\delta Q_i](s) = \im\sum_{KL}w_L\,q_i^{KL}\frac{\Laplace[a_{LK}](s)}{s - \im\Omega_{KL}} + \text{c.c.} \, ,
\end{align*}
where the Laplace transform is defined as
\begin{align*}
\Laplace [f](s) \isDefinedAs \binteg{t}{0}{\infty} \e^{-st}f(t) .
\end{align*}
To derive our working condition, we multiply by the Laplace transform of the perturbing potentials, $\Laplace[\delta v_i](s)$ and sum over the remaining index
\begin{align*}
\sum_i\Laplace[\delta v_i](s)\,\Laplace[\delta Q_i](s)
= -2\sum_{KL}w_L\,\frac{\Omega_{KL}}{s^2 + \Omega_{KL}^2}\abs{\Laplace[a_{LK}](s)}^2 \, .
\end{align*}
Hence, in absence of response we find the following necessary condition
\begin{align*}
0 = \sum_{KL}\frac{w_L\Omega_{KL}}{s^2 + \Omega_{KL}^2}\abs{\Laplace[a_{KL}](s)}^2 \, .
\end{align*}
Note that each $KL$ pair occurs twice in this summation and the terms $K = L$ do not contribute. The necessary condition can therefore be rewritten as
\begin{align}\label{eq:nesCond}
0 = \sum_{\crampedclap{K > L}}\frac{(w_L - w_K)\Omega_{KL}}{s^2 + \Omega_{KL}^2}\abs{\Laplace[a_{KL}](s)}^2 \, .
\end{align}
Similar as in the proof for ground states, we would like to infer that this condition implies that all individual terms need to be zero. To do so, we demand that all individual terms are positive or zero, so we impose the following condition on the ensemble
\begin{align*}
(w_L - w_K)\Omega_{KL} \geq 0 \qquad \text{for all $K$ and $L$} \, .
\end{align*}
So that means that for any two states with different energies with $E_K < E_L$, we require $w_K \geq w_L$. This condition does not yield additional restrictions for degenerate states.\footnote{Also $(w_L - w_K)\Omega_{KL} \leq 0$ would work as a condition on the ensemble. This option  works only for Hamiltonians bounded form above (finite basis), since the highest weight is associated to the highest energy. Further, this option is physically not very sensible, so this case is not explicitly pursued.} Note that the equality is not only satisfied for degenerate states, but also states with equal weights and possibly different energies. States for which the equality holds need a special treatment, so for each state we define
\begin{align*}
\mathcal{D}(K) \isDefinedAs \bigl\{L : (w_L - w_K)\Omega_{KL} = 0\bigr\}
\equiv \bigl\{L : E_L = E_K \lor w_L = w_K\bigr\} \, .
\end{align*}
I will refer to this subspace as the `extended degenerate subspace' of the state $K$, since not only states with the same energy belong to it, but also states with the same weight.

Armed with this additional condition on the ensemble let us return to the necessary condition~\eqref{eq:nesCond}. Since all the individual terms are zero or positive, we find that $\Laplace[a_{KL}](s) = 0$ if $L \notin \mathcal{D}(K)$. So in the time-domain we find that $a_{KL}(t) = 0$ almost everywhere. The first possibility is that $\delta v_j(t) = 0$ almost everywhere. Typically one would also require $\delta v_j(t) \in C^1[0,T]$ (differentiable up to first order in the time-interval $[0,T]$) to guarantee a physical (strong) solution,\footnote{In Ref.~\cite{RuggenthalerPenzLeeuwen2015} it is stated that the condition $\delta v_j(t) \in C^1$ can probably be weakened to Lipschitz continuity. This is still sufficient for our argument, since we only need continuity. A milder version of the Schrödinger equation would allow for more general potentials in some $L^p$ spaces in time~\cite{PenzRuggenthaler2015, RuggenthalerPenzLeeuwen2015}. In that case, however, potentials which only differ at a set of zero measure would be considered equivalent.} so the term `almost everywhere' is of no importance in physical situations. For more details on the solvability of the time-dependent Schrödinger equation and in particular when physical (strong) solutions exist, I refer the reader to an excellent exposition by Ruggenthaler, Penz and Van Leeuwen~\cite{RuggenthalerPenzLeeuwen2015}. This first possibility is the trivial way to have no response (no potential), so is not the solution we are interested in.

The more interesting possibility is the existence of linear combinations of the operator $\hat{Q}_j$, 
\begin{align*}
\hat{L}_n = \sum_j\hat{Q}_j\delta v_j^n \, ,
\end{align*}
such that $l_n^{LK} = \brakket{\Psi_L}{\hat{L}_n}{\Psi_K} = 0$ for all pairs $K, L \notin \mathcal{D}(K)$. In other words, we should look for operators $\hat{L}_n$ such that they only give components in the extended degenerate subspace
\begin{align}\label{eq:necessaryCondition}
\hat{L}_n\ket{\Psi_K} = \sum_{\crampedclap{L \in \mathcal{D}(K)}}l_n^{LK}\ket{\Psi_L} \, .
\end{align}
If any of such linear combinations exist, they are candidates to belong to the kernel of the retarded response function.

\section{Sufficiency}
\label{sec:sufCond}
In the previous section we have derived a necessary condition~\eqref{eq:necessaryCondition} for potentials in the kernel of the response function. Suppose that we have found such a potential, we will now check to which extend this condition~\eqref{eq:necessaryCondition} is also sufficient. We start by rewriting the Lehmann representation of the response function~\eqref{eq:Lehmann} by retaining only the unique pairs in the sum
\begin{align*}
\chi_{ij}(t-t') 
&= \im\theta(t-t')\sum_{K > L}(w_L - w_K)\,\e^{\im\Omega_{KL}(t-t')}q_j^{LK}q_i^{KL} + \text{c.c.} \, .
\end{align*}
I used here that the $K = L$ terms drop out of the summation. 
If there exists one or more operators $\hat{L}_n$ which satisfy~\eqref{eq:necessaryCondition}, we find that the response is only truly zero if
\begin{align*}
0 = \im\sum_{\mathclap{L < K \in \mathcal{D}(L)}}(w_L - w_K)\,\e^{\im\Omega_{KL}(t-t')}q_j^{LK}l_n^{KL} + \text{c.c.} 
\quad \forall_j \, .
\end{align*}
Note that due to necessary condition~\eqref{eq:necessaryCondition}, the sum could be restricted to pairs in their respective extended degenerate subspace.
Since states with equal weights do not actually contribute to this sum, we can further restrict the summation over pairs which have different weights, so only (energetically) degenerate states can contribute
\begin{align}\label{eq:suffiencyCondition}
0 = \im\sum_{\mathclap{L < K \in \mathcal{D}^r(L)}}(w_L - w_K)q_j^{LK}l_n^{KL} + \text{c.c.} 
\quad \forall_j \, .
\end{align}
where we used that $\Omega_{KL} = 0$ for degenerate states and introduced
\begin{align*}
\mathcal{D}^r(L) \isDefinedAs \bigl\{K : E_K = E_L \land w_K \neq w_L\bigr\} \, .
\end{align*}
From this form of the sufficiency condition~\eqref{eq:suffiencyCondition} we can extract two important results. The first result is a generalization of the commutator form for the sufficiency condition~\cite{Giesbertz2015}. Since for other pairs $L$ and $K \notin \mathcal{D}^r(L)$ we have that either $w_L = w_K$ or $l_n^{KL} = 0$, we can put them back into the summation without affecting the result. This yields the following form for the sufficiency condition
\begin{align}
0 &= \sum_{\mathclap{L,K}}(w_L - w_K)q_i^{LK}l_n^{KL} + \text{c.c.} \notag \\
&= \sum_Kw_K\brakket[\big]{\Psi_K}{\bigl[\hat{Q}_i,\hat{L}_n\bigr]}{\Psi_K} = 0 \qquad \forall_i \, .
\end{align}%
Note that this expression correctly reduces to the pure state case if $w_0 = 1$ (so $w_{i>0} = 0$). It will be hard to check this condition for general ensembles.

A more useful result is obtained by noting that if the weights only depend on the energy, $w_K = w(E_K)$, that for degenerate state the weights are equal. This implies that $D^r(K) = \emptyset$ for any $K$, so the sufficiency condition~\eqref{eq:suffiencyCondition} is trivially satisfied, since the sum does not run over anything anymore. Hence, condition~\eqref{eq:necessaryCondition} is for these ensembles not only necessary, but also sufficient. Note that the canonical and the grand canonical ensembles exactly belong this category of ensembles.

\section{(Grand) Canonical ensembles}
\label{sec:thermoEnsembles}
In this section we will specialize ourselves to the (grand) canonical ensembles.
In the (grand) canonical ensemble all states in the Hilbert space contribute. The only difference between the canonical and grand canonical ensemble is the extend of their Hilbert spaces. In the canonical ensemble the Hilbert space only consists of states with a specific particle number, $N$. In the grand canonical ensemble also states with a different particle number are included, so the full Fock space is taken as the Hilbert space. The number of particles is then regulated via a constant shift in the potential: the chemical potential.

The most important thing to realise is that all states contribute to response function in the (grand) canonical ensemble, because all weights are non-zero. So the necessary and sufficient condition~\eqref{eq:necessaryCondition} can only be satisfied if and only if all eigenstates of the unperturbed Hamiltonian $H_0$ can be expressed as eigenstates of the operator $\hat{L}_n$. This is only the case if the operator $\hat{L}_n$ commutes with the Hamiltonian.
\begin{align}\label{eq:commuteCond}
\bigl[\hat{H}_0,\hat{L}_n\bigr] = 0 \, .
\end{align}
Note that this is a much more stringent (and convenient) condition than in the $T=0$ case~\cite{Giesbertz2015}. In the non-degenerate pure state case, only the ground state needs to be an eigenstate of the operator $\hat{L}_n$ and the excited states are immaterial. In the degenerate $T = 0$ case, the operator $\hat{L}_n$ is allowed to create components in the degenerate subspace, but again, the excited states do not play any role.

In the case of the retarded 1RDM response function, the weaker condition in the $T = 0$ case does give rise to some `pathological' potentials which do not commute with the initial Hamiltonian, $\hat{H}_0$, as described in the introduction.
In the (grand) canonical ensemble case, however, only the potentials related to the symmetries of the Hamiltonian remain due to condition~\eqref{eq:commuteCond}. In other words, \emph{all} eigenstates of the initial Hamiltonian should be choosable as eigenfunctions of the operator $\hat{L}_n$. Hence, the potentials in the kernel of the 1RDM response function for the (grand) canonical ensemble exactly coincide with the symmetry-induced ones of the zero temperature 1RDM response function. Some examples:
\begin{itemize}
\item Number conserving Hamiltonian: $\hat{N}$.
\item Spin-independent Hamiltonian: $\hat{\mat{S}}$.
\item Linear molecule: $\hat{L}_z$.
\item (Spherical) atoms: $\hat{\mat{L}}$.
\item Homogeneous electron gas: $\hat{\mat{L}}$ and $\hat{\mat{p}} \isDefinedAs -\im\nabla$.
\end{itemize}
There does not seem to be any difference between the kernels of the 1RDM response function in the canonical and grand canonical ensemble. The only case I can think of is the one-particle case in combination with an interacting Hamiltonian. In the canonical ensemble, only the 1-particle sector of the Fock space is used, so effectively the Hamiltonian is non-interacting. Since there is only one particle, \emph{all} symmetry operators can be expressed as one-body operators even the discrete ones. In the grand canonical ensemble, also the states with a higher number of particles contribute. For these states, however, the discrete symmetries cannot be expressed as one-body operators. Hence, the discrete symmetries of the system are part of the kernel of the 1RDM response function in the $N=1$ case, only in the canonical ensemble and not in the grand canonical ensemble. Such a situation where the kernel of the retarded 1RDM response differs between the canonical and grand canonical ensembles, seems only to appear when $N = 1$.

There is one exceptional case I would like to mention in this context, where the broken symmetry due to the interaction is actually a continuous one. This is the so-called Runge--Lenz vector in hydrogenic systems
\begin{align*}
\half\bigl(\hat{\mat{p}} \wedge \hat{\mat{L}} - \hat{\mat{L}} \wedge \hat{\mat{p}}\bigr) - Z\frac{\vecr}{\abs{\vecr}} ,
\end{align*}
where $Z$ is the atomic number.
The Runge--Lenz vector explains the degeneracy between the hydrogenic orbitals with different angular momenta, but the same principle quantum number. So in hydrogenic systems the Runge--Lenz vector will be part of the kernel of the 1RDM response function, if a canonical ensemble is considered as initial state. The Runge--Lenz vector, however, is only a symmetry for one-electron states, because the electron-electron interaction breaks this symmetry (the degeneracy between the $2s$ and $2p$ state is lifted). Since the macro canonical ensemble also includes states with more than one electron, the Runge--Lenz vector will therefore not be part of the kernel of the 1RDM response function for the grand canonical ensemble.

For completeness, I would like to point out that the same analysis implies that in the case of the density response function, nothing changes when going from the zero temperature to the finite temperature formalism. The only potential in the kernel of the density response function at $T = 0$ is the constant potential~\cite{Leeuwen2001, Giesbertz2015} for particle number conserving Hamiltonians, $\bigl[\hat{H}_0,\hat{N}\bigr] = 0$. Since this potential corresponds to a symmetry of the Hamiltonian, this potential will also be the only possible component in the kernel of the density response function at $T > 0$. Hence also at finite temperature the density response function is invertible up to a constant for particle conserving Hamiltonians. The same result has also been obtained in a less general derivation limited only to the finite temperature density response function~\cite{Pribram-JonesGrabowskiBurke2015}.

\section{What about time-independent response functions?}
\label{sec:timeINdependent}
In Ref.~\cite{Giesbertz2015} I described how the $T = 0$ results for the time-dependent response function carried over to the time-independent response function for the non-degenerate ground state case. In the ensemble case such a transfer is not possible. The reason is that in the time-dependent response function, only the states are affected by the perturbations and not the weights of the ensemble as reflected by the commutator condition~\eqref{eq:commuteCond}. So only the initial ensemble is in a thermodynamic equilibrium with the bath and at later times the system is essentially uncoupled from the bath.

On the contrary, in the time-independent case also perturbations in the weights are taken into account, via their dependence on the energies. Hence, in the time-independent case the response function does not only depend on the perturbation of the eigenstates, but also on the perturbation of the eigenvalues. This additional dependence therefore eliminates all operators from the kernel of the time-independent 1RDM response function for the (grand) canonical ensemble, since the both the eigenstates and eigenvalues need to be unperturbed. Only the number operator would remain for the canonical ensemble, since all states in the corresponding Hilbert space are degenerate with respect to the number operator, i.e.\ the constant shift (chemical potential) does not affect the canonical ensemble. This result is exactly in agreement with previous work on finite temperature 1RDM functional theory by Van Leeuwen~\cite{Leeuwen2007} and Baldsiefen~\cite{BaldsiefenGross2012a, PhD-Baldsiefen2012}.

\section{Conclusion}
I have further generalized the invertibility proof by van Leeuwen for the density response function~\cite{Leeuwen2001}. This generalized proof is not only valid for the ground state as initial state, but works for general ensemble in which the weights of higher lying states are smaller than the weights of lower lying states. Two conditions were derived which are necessary and sufficient for a perturbation to yield no response at all. Without additional assumptions on the structure of the ensemble, these conditions are too cumbersome to make general statements. One of these conditions could be eliminated altogether, by requiring that the value of the weights are directly related to the energies of the states. Demanding additionally that all states participate in the ensemble (have a weight strictly larger than zero), the remaining condition simplifies even further to the requirement that potentials can be part of the kernel of the response if and only if they commute with the initial Hamiltonian. In particular for thermodynamic ensembles these assumptions hold.

That the potentials in the kernel of the response function need to commute with the Hamiltonian for thermodynamic ensembles is a much more stringent condition on these potentials than in the pure state case. This is a significant advantage when investigating the invertibility of the 1RDM response function. In the pure state case (zero temperature) one needs to take many `pathological' potentials into consideration which are related to the structure of the ground state. In finite temperature only potentials related to the symmetries of the system remain and these `pathological' potentials are of no concern anymore. Since such `pathological' potentials were not present in the density response function at zero temperature, the kernel of the density response function is the same in the zero and finite temperature formalism.

\section*{Acknowledgements}
The author would like to thank prof.dr.\ R.~van Leeuwen for stimulating discussions.
Support from the Netherlands Foundation for Research NWO (722.012.013) through a VENI grant is gratefully acknowledged.

\bibliographystyle{tMPH}
\bibliography{bible}

\end{document}